\documentclass[useAMS,usenatbib]{mn2e} 
 
\usepackage{mathrsfs} 
\usepackage{graphicx} 
\usepackage{mediabb}
\usepackage{color}  

\def\red{\color{red}}  
\def\red{ }  
\def\exp{{\rm exp}} \def\sech{{\rm sech}}
\def\kms{km s$^{-1}$} \def\Kkms{K \kms}

    \def\Msun{M_\odot} 
\def\deg{^\circ} \def\Tb{T_{\rm B}}

\def\Hcc{ H cm$^{-3}$ }\def\mH{m_{\rm H}}\def\mHcc{$\mH \ {\rm cm}^{-3}$}
\def\cos{{\rm cos}} \def\/{\over} 
 
\def\ad{a_{\rm d}}\def\rdisk{a_{\rm d}}
\def\kpc{{\rm kpc}}\def\A{\mathcal{A}}\def\sech{{\rm sech}}
\def\({\left(} \def\){\right)} \def\[{\left[} \def\]{\right]}
\def\be{\begin{equation}} \def\ee{\end{equation}}
\def\g{\gamma}\def\gm{(\gamma-1)} \def\gp{(\gamma+1)}
\def\J{J(R,\theta)} \def\I{I(R,\theta)}
\def\rdisk{a_{\rm d}} \def\({\left(} \def\){\right)}
\def\kpc{{\rm kpc}}\def\B{\mathcal{B}}\def\C{\mathcal{C}}\def\S{\mathcal{S}}

\def\r{\varpi}
\def\U{\mathcal{\delta U}}

\def\G{g}
\def\d{\partial} 
\def\Vrot{V_{\rm rot}} 
\def\Vav{\langle V_{\rm AM} \rangle}

\begin{document}    
\title[{  Halo Diagnostics by Galactic-Centre Shock Wave}]
{ {
Diagnostics of Gaseous Halo of the Milky Way by a Shock Wave from the Galactic Centre}}    
\author[Y. Sofue 
]
{Yoshiaki {\sc Sofue}$^{1}$\thanks{E-mail: sofue@ioa.s.u-tokyo.ac.jp} 
\\
1. Institute of Astronomy, The University of Tokyo, Mitaka, Tokyo 181-0015, Japan  
}
   
\maketitle 
 
\begin{abstract}  
A method to diagnose the gas distribution in the Galactic halo using a shock wave from the Galactic Centre (GC) is presented. Propagation of a shock wave caused by central explosion with released energy $\sim 10^{55}$ erg is calculated for various models of the gaseous halo, and the shock front morphology is compared with the observed North- (NPS) and South-Polar Spurs (SPS). The observed bipolar hyper-shell (BHS) structure of the spurs is reproduced, when a semi-exponential halo model with radially-variable scale height is adopted. A spherical halo model ($\beta$ model) is shown to be not appropriate to explain the observed BHS shape. Asymmetry of the spurs with respect to the rotation axis and to the galactic plane is explained by large-scale density gradient in the halo across the Galaxy. Such gradient may be produced by ram pressure of the IGM during motion of the Galaxy in the Local Group. A halo model having sinusoidal density fluctuations (clouds) can also explain the asymmetry, given appropriate cloud parameters are chosen. Further irregular features such as filaments and multiple curvatures superposed on the spurs are also understood as due to density fluctuations in the halo. 
\end{abstract}

 
\section{Introduction} 

The origin of the North Polar Spur (NPS) and its southern counter spurs (SPS) have been explained by a shock wave from the Galactic Centre (GC) driven by an explosive event $\sim 10$ My ago with released energy $\sim 10^{55-56}$ erg such as due to starburst or nucleus explosion (Sofue 1977, 1984, 1994, 2000; Sofue et al. 2017; Sarkar et al. 2015, 2016; Kataoka et al. 2018 for a review). High-energy explosive events at the GC have been confirmed by the discovery of Fermi Bubbles in $\gamma$ rays (Su et al. 2010). The X- and $\gamma$-ray bubbles are considered to have common origin at different epochs (Crocker et al. 2015; Kataoka et al. 2018).
{It has been suggested that a shock wave from the GC can be used to diagnose the distribution of gas in the Galactic halo by comparing the calculated shock front with the observed spurs (Sofue 1994, 2000). 

The existence of a gaseous halo surrounding the Galaxy has been a decades of discussion in relation to the missing baryon problem in the Local Group (Fang et al 2013 and the literature therein).  The spatial distribution of the gas is a clue to discriminate possible origin of the hot halo, namely, a disk-like halo would prefer a galactic-disk origin, while a spherical halo would favor inflow origin from the intergalactic space. 

High temperature gaseous halo around the galactic disk has been detected by UV (Sembach and Savage 1992; Shull and Slavin 1994) and soft X-ray (Miller and Bregman 2013, 2015, 2016; Nakashima et al. 2018; and the literature therein). Recent analyses of soft-X ray observations on board SUZAKU have revealed further details of the density and temperature structures in the Galactic halo (Kataoka et al. 2017; Nakashima et al. 2018; Akita et al. 2018). A disk model has been shown to better fit the observed X-ray spectroscopic observations than a spherical model (Nakashima et al. 2018). Anisotropy in the column density of hot gas has been detected in several restricted directions, indicating that the halo gas is inhomogeneous (Gupta et al. 2014; 2017). 
{\red Effects of the asymmetry and/or wind in the halo on the shock wave propagation have been also studied in relation to the observed asymmetry of the NPS and Fermi Bubbles (Sofue 1994, 2000; Mou et al. 2018; Sarkar 2019).}

In the present paper, we discuss the possibility to use the NPS and SPS to diagnose the gaseous halo by calculating the propagation of a shock wave from the Galactic Centre assuming a variety of types of model halos. Thereby, an advanced halo model of the Galaxy based on the recent X-ray observations will be constructed as the basic, standard halo model. The standard model will be added of perturbations and/or modification to examine how the shock front shape depends on the halo models. By comparing the calculated results with the observed spur's morphology, qualitative diagonostics will be obtained of the gaseous distribution in the Galactic halo. 
}

\begin{figure*} 
\begin{center}      
\includegraphics[width=15cm]{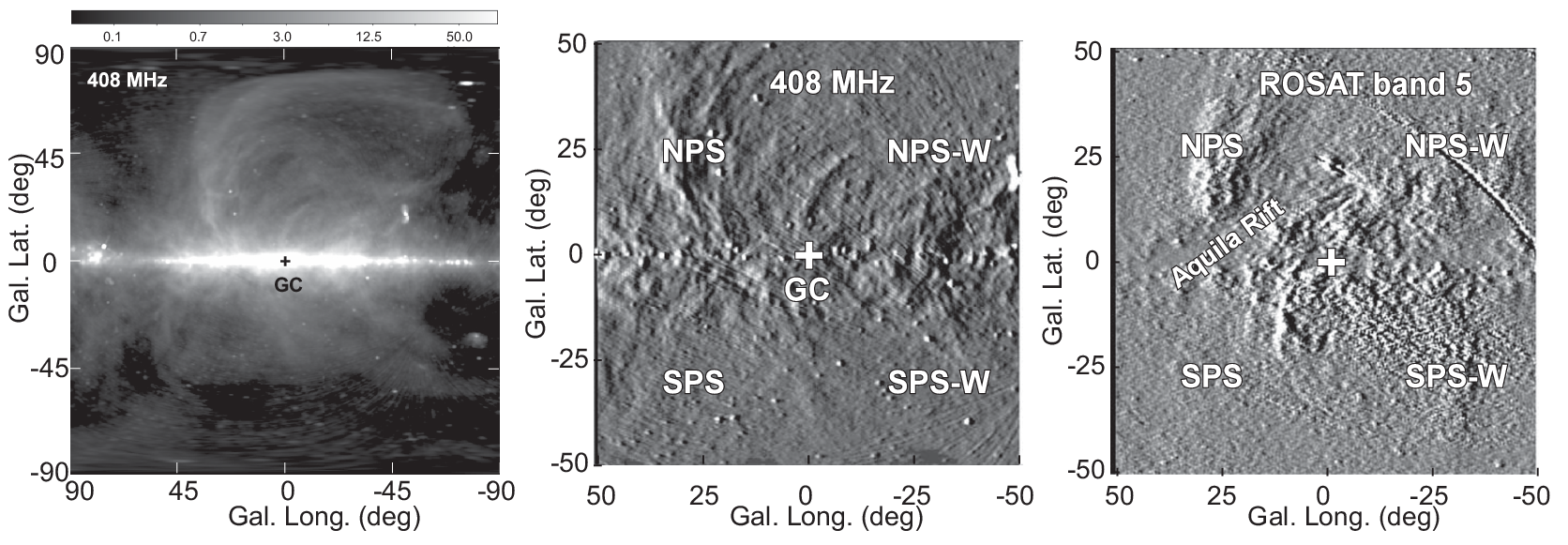} 
\end{center}
\caption{{[Left] 408 MHz radio continuum intensity map taken from the Bonn-Parkes all sky survey (Halsma et al. 1982). Intensity scale bar in K of $T_{\rm b}$.}  [Middle] Relief map of the 408 MHz map taken from Sofue (2000). [Right] Same for ROSAT band 5 map ($\sim 1$ keV) using the data from Snowden et al. (1993).}
\label{relief}  
\end{figure*}

\section{Galactic-Centre Spurs} 

Observations of radio continuum and soft X-ray structures of the NPS and related spurs around the GC are over-viewed in this section.
{
Figure \ref{relief} shows a 408 MHz radio continuum map of the $180\deg \times 180\deg$ region around the GC and a relief image for $100\deg \times 100\deg$ region made from the all-sky survey by Haslam et al. (1982). }
The relief method was also applied to the 1408 MHz (Reich et al. 2001) and 2300 MHz (Jonas et al. 1985) all sky maps, which confirmed that the detected spurs in figure \ref{relief} are recognized to be mostly real. Also shown in the figure is a relief image of the ROSAT band 5 (1 keV) soft X-ray map using the data from Snowden et al. (1993). 

The relief technique was applied in order enhance the spurs, with which extended background emissions are subtracted and small-scale structures embedded in the bright disk are enhanced. The reliving was so applied that the image was shifted in the longitude direction by about the angular resolution, and was subtracted from the original. By this procedure, the strong galactic background is subtracted and small scale structures are enhanced. Shifting in the longitude direction makes the maps more sensitive to vertical structures. Note, however, that the relieved images may not be used for quantitative measurement of the intensity, but is useful only for morphological study of edged structures. Three radio maps at different frequencies from independent observations have been structured in order to confirm that the enhanced features are mostly real. Note that some slight artificial stripes remain in some portions of the maps, which are due to scanning effects during the observations. 
 
The NPS is visible as the strong radio and X-ray ridge at $\sim$G30+20, which extends toward the galactic plane and merges into the complex disk emission at $\sim$G21. The western counterpart to NPS is recognized at $\sim$G-30+20 extending from the galactic plane to positive latitude, which was called the NPS-West (Sofue 2000). Southern counterparts are recognized at $\sim$G30-20 and $\sim$G-30-20, which are called the South Polar Spur (SPS) and West (SPS-W). 

The major spurs are also recognized in the relieved X-ray map in figure \ref{relief}. The X-ray NPS is clearly visible coincident with, while slightly inside, the radio NPS. The radio SPS-W is associated with an X-ray spur at $\sim$G-30-20, drawing a clear arc symmetric to the NPS about the GC. The radio NPS-W is also associated with an X-ray spur at G30+20, and SPS-W is also associated with an X-ray spur at $\sim$G30-30, while both are fainter than the other three spurs. 

{
These four spurs in radio and X rays (NPS, NPS-W, SPS and SPS-W) compose a bipolar double-horn structure around the GC, which we call the bipolar hyper shell (BHS). The whole BHS is in general symmetric around the GC, while some deformation and asymmetries with respect to the rotation axis as well as to the galactic plane are recognized. The morphology will be used for diagnosing the gaseous halo by comparing with simulated shock front shapes.
}

The NPS has been often called Loop I drawn on the sky of diameter $120\deg$ cantered on G$-35+18\ (l=-35\deg,\ b=+18\deg)$ with the western arc crossing the galactic plane at G$-90+00$ (Berkhuijsen et al. 1971; Shchekinov 2018). However, difficulties have been pointed out on the Loop hypothesis as follows. 
(i) Radio emission is observed only in the NPS, but missing in the other three quarters of the loop. 
(ii) The major radio and X-ray spurs (NPS-W, SPS, and SPS-W) in the central $100\deg\times 100\deg$ around the GC (figure \ref{relief}) are ignored.  
(iii) The idea assumes four Loops (I to IV) on the whole sky at distances $\sim 100-300$ pc, which requires an exceptionally high rate of supernova remnants near the Sun.
(iv) The distance to NPS has been determined to be greater than a few kpc. Hence, the loop requires a sphere of radius $>\sim 10$ kpc crossing the Galaxy. So, in this paper, the GC origin hypothesis is adopted for the major spurs. 
{
(v) No H$\alpha$ and optical line emissions, typical for supernova remnants, are detected even in the strongest ridge of the NPS, indicating that the ambient gas is hotter than $\sim 10^6$ K.
  }

\section{Shock Wave Propagation} 

\subsection{Bipolar hyper shell model}

The double horn structure composed of the NPS, NPS-W, SPS and SPS-W has been simulated by the bipolar-hyper shell (BHS) model (Sofue 1977, 1984, 1994, 2000, 2017; Sofue et al. 2016; Sarkar et al. 2016 ). Although the global morphology of the spurs is understood by these axisymmetric models, further models may be also worth to explain the asymmetric morphology of the spurs as well. 

The GC explosion is assumed have taken place $\sim 15$ My ago with released energy of $E\sim  10^{55}$, which is required to explain the observed temperature of the NPS by soft X-ray measurements. 
The shock wave propagation is calculated by solving the second order derivative equation of the shock front radius $R$ with respect to the elapsed time $t$ since the explosion at the GC using the Sakashita's (1971) method (see also Sofue 1984, 2000, 2017; M\"oellenhoff 1976 ). 
{The method has been improved by including the gravitational force.  The detail of the formulation is given in Appendix A.}

The shock front at Galacto-centric distance, $R$, is assumed to propagate radially in the adiabatic gas. The swept-up gas from the interior of shock wave is accumulated in the thin surface of the front. The released energy at GC is accumulated in each solid angle shell as thermal and kinetic energies of expanding motion. So, the density and temperature variations cannot be discussed in this method.  

\subsection{Models of Unperturbed Gaseous Halo and Disk}  
\label{densitymodel}

The unperturbed gas distribution in the galactic disk and halo is represented by superposition of several gaseous layers and uniform intergalactic gas in the cylindrical coordinate system as
$${  \rho(\r,z) =\Sigma_i \rho_i(\r,z)} $$
\be=\rho_{\rm CMZ}+\rho_{\rm mol}+\rho_{\rm HI}+\rho_{\rm wHI}+\rho_{\rm UV}
+ \rho_{\rm X}+\rho_{\rm IGM}.
\ee
{ 
Here, $\r$ and $z$ are the distance from the rotation axis and height from the galactic plane, respectively, in the cylindrical coordinates, which are related to the galacto-centric distance $r$ and elevation angle $\theta$ through $\r=r\ \cos \ \theta$ and $z=\r\ \sin \ \theta$. Detailed formulation is given in Appendix \ref{appdensitymodel}.
}

As a standard model, we consider the following components: (1) the central molecular zone (CMZ), (2) main molecular disk, (3) cold HI disk, (4) warm HI halo (disk), (5) hot ionized (UV) halo (or thick disk) of temperature $T\sim 10^5$ K, (6) hot X-ray halo of $T\sim 10^{6-7}$ K, and (7) uniform intergalactic gas (IGM). 

For the disk-like components, the vertical scale heights, $z_i$, are assumed to increase with the distance from the rotation axis, $\r$, obeying the semi-exponential form as described in Appendix \ref{appdensitymodel} (inversely to figure  \ref{SemiExpSech}). The increasing scale heights manifest varying vertical potential that is deeper in the center.  

\begin{figure}
\begin{center}   
\includegraphics[width=6cm]{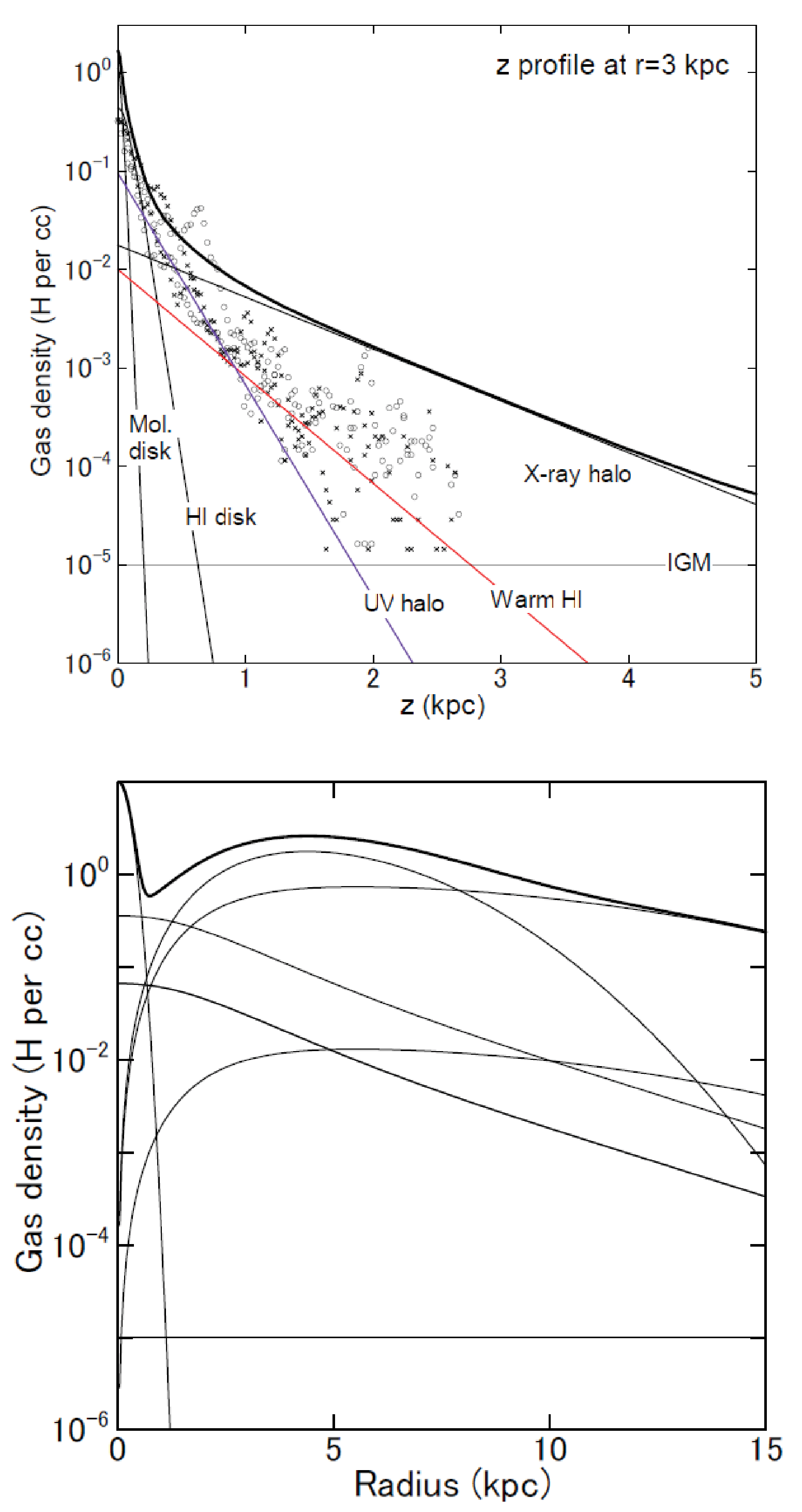} 
\end{center}
\caption{[Top] Vertical gas density profiles at $r \sim 3$ kpc of the molecular disk, HI disk, hot UV halo of $T\sim 10^5$ K, X-ray halo of $ 10^{6-7}$ K, and intergalactic matter. Dots are measured HI densities at $l \sim 22\deg$ and $V_{\rm lsr}\sim 120$ \kms using the HI survey data by McClure-Griffiths et al. (2009).  
[Bottom] Radial density profile in the galactic plane.   }
 \label{rhoprof}  
\end{figure}

Figure \ref{rhoprof} shows the density profiles across the galactic plane at $l=22\deg$ and in the galactic plane. Two dimensional density maps are shown in the following subsection  along with the calculated results.

We further modify this standard model by adding various perturbations in order to examine the effect of local and global density fluctuations on the morphological evolution of the shock front.

\subsection{Shock Propagation in the standard model halo}
\label{shockstandard} 

We numerically solve the Sakashita's equation (\ref{eqsakashita}) by integrating the second derivative of $R$, $\ddot{R}$, with respect to the elapsed time from the explosion, $t$, for a given gas density distribution, $\rho(r,\theta)$, as described in subsection \ref{densitymodel}. 
{  
Here, $R$ is the galacto-centric radius of the shock front, $r$ is the radius to describe the unperturbed gas distribution, and $\theta$ is the elevataion angle from the galactic plane.
}
{Thereby, we modified the equation by introducing the Galactic gravitational force. See Appendix for the detail of the Sakashita's method. }

\begin{figure} 
\begin{center}      
\includegraphics[width=6cm]{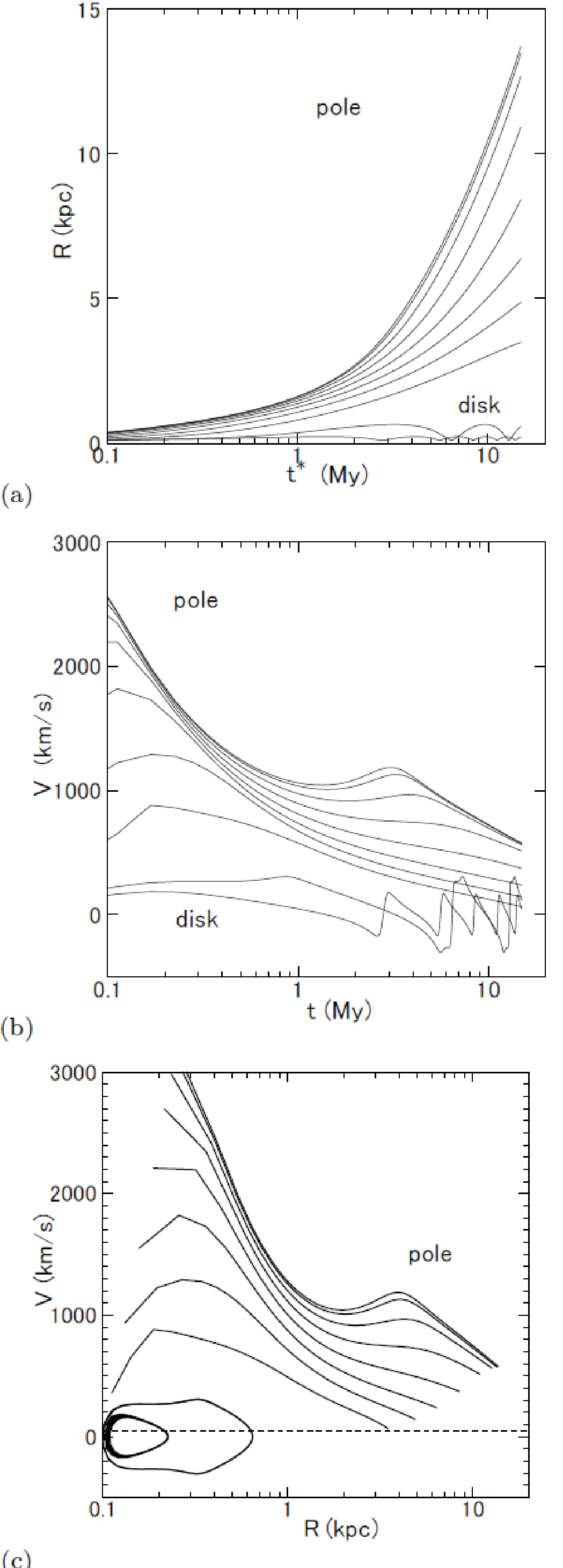} 
\end{center}
\caption{(a) Time evolution of the radius, (b) velocity, and (c) velocity against radius. }
\label{timeevo}  
\end{figure}

\begin{figure*} 
\begin{center} 
\includegraphics[width=13cm]{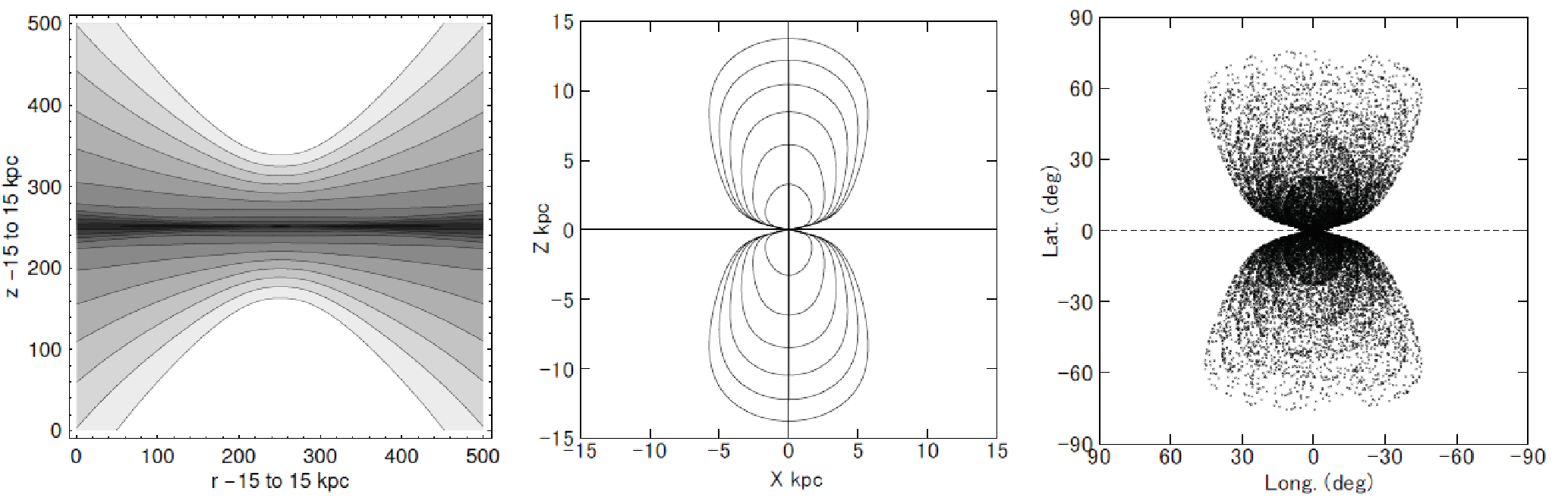}  
\end{center}
\caption{Meridional density map of the disk and halo gases within $\pm 15$ kpc from the GC for the standard (semi-exponential) halo model (left). Contours are drawn every $\sqrt{10}$ times the neighbors starting from $10^{-5}$ \Hcc (white region) and ending at $10$ \Hcc (darkest region), shock wave front from t=2.5 to 20 My in the $(X,Z)$ plane plotted every 2.5 My for $E=E_0= 1.5\times 10^{55}$ erg (middle), and the front projected on the sky every 5 My (right). Same plots apply for different $E$ by changing the time to $t=t_{\rm plot}(E/E_0)^{-1/2}$.
}
 \label{model}  
 \label{rhomodel}
\begin{center} 
\includegraphics[width=13cm]{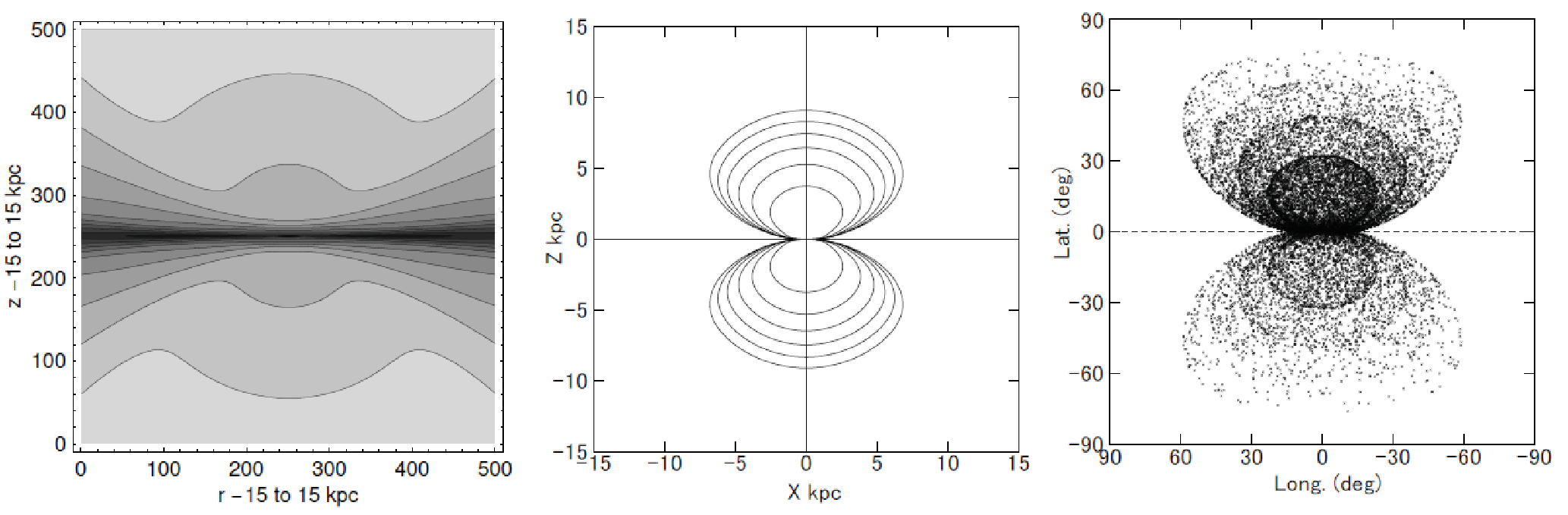}  
\end{center}
\caption{Meridional density map of the spherical ($\beta$ model) halo (left), shock front from t=2.5 to 20 My in the $(X,Z)$ plane plotted every 2.5 My for $E=E_0= 1.5\times 10^{55}$ erg (middle), and the front projected on the sky every 5 My (right).  }
 \label{model-sph}  
\end{figure*}

 Calculated result for the standard halo model is shown in figures \ref{timeevo} and \ref{model}.
Figure \ref{timeevo} shows time evolution of front radius $R$ and expansion velocity $V=\dot{R}$ for $E= 1.5\times 10^{55}$ erg for ray paths at different elevation angles from $\theta=0\deg$ (galactic plane) to 90$\deg$ (north galactic pole). Because the initial explosion is assumed to start on a finite sphere of radius 0.1 kpc, ray paths mostly start outside the molecular disk of thickness 0.02 kpc except for $\theta<\sim 10\deg$. Time for different energy $E$ is obtained by $t=t_{\rm plot}(E/E_0)^{-1/2}$ My and velocity by $V=V_{\rm plot} \times (E/E_0)^{1/2}$, where $E_0= 1.5\times 10^{55}$ erg. Also, the input energy $E$ may be changed according to the appropriate change of the the time scale unit. Namely, for different $E$ value, the same plots apply by changing the time by a factor of $t/t_{\rm calc.}= (E/E_0)^{-1/2}$.  

Figure \ref{model} shows a meridional cross section of the shock front projected on the $(X,Z)$ plane every 2 My for $E= 1.5\times 10^{55}$ erg. As the nature of the Sedov type solution, the front shape is identical for the same value of $E/\rho_0$, so that the plotted result applies to any combination of $E$ and $\rho_0$ having the same ratio. 

The front expansion is decelerated in the disk at $R<1$ kpc. When the rays escape from the disk and enter the halo, the rays are again decelerated by the extended halo gas as well as by the constant-density intergalactic gas. The wavy behavior of the expansion velocity is due to interchange from disk to halo, and from halo to IGM.

{
In the central region, while most of the released energy escapes into the halo, making the bipolar dumbbell-shaped front, the shock wave near the galactic plane at $\theta < \sim 10\deg$, is strongly decelerated by the dense disk.
The CMZ, as well as the inner CO and HI disk, is considered to be destroyed by the initial fast expansion of the shock wave. However, the once blown-off gas in the disk is, then, rapidly attracted back toward the GC due to steep gradient of the gravitational potential of the central bulge within a time scale of a few My. The presently observed CMZ may be a remnant disk, which had been destroyed $\sim 10$ My ago, and reconstructed soon after the explosion. 

It is interesting to note that the front radius oscillates inside the disk (figure \ref{timeevo}), which occurs due to exchange of the conserved internal gaseous energy (pressure), kinetic and gravitational energy under the adiabatic condition. However, in a more realistic ISM, such oscillation will be damped down quickly due to dissipation of the shock wave as well as due to energy loss by radiative cooling. 
}

As shown in figure \ref{model}, the front expands into the low-density halo to form a bipolar dumbbell shape. The front is strongly bent near the galactic plane due to the deceleration by the dense and heavy molecular disk, so that the dumbbell shape becomes sharply kinked in the galactic plane. The shock front for $E \sim 1.5\times  10^{55}$ erg at elapsed time $t \sim 15$ My well resembles the NPS.  

\subsection{X-ray temperature and shock wave velocity}

The electron temperature of the NPS has been measured at 16 ON positions along the X-ray ridge at latitudes $b=43\deg$ to $74\deg$ by observing soft X-ray spectra with the SUZAKU satellite (Akita et al. 2018). The averaged temperature of the 16 measurements ON the NPS is calculated to be $kT=0.33\pm 0.07$ keV ($T=3.8\times 10^6$ K), or sound velocity $c_{\rm s}=\sqrt{\gamma k T/\mH}=230$ \kms, which requires for the shock front velocity greater than $\sim 230$ \kms. This temperature may be compared with the temperature obtained for the galactic halo of $kT=0.26$ keV ($3.0\times 10^6$ K) outside NPS (Nakashima et al. 2018), indicating that the NPS is a supersonic phenomenon in the halo.

{
\section{Diagnostics of Gaseous Halo}
}

The result for the standard halo model could reasonably reproduce the global morphology of the observed spurs. In this section we examine several different symmetric models in order to confirm that the standard model is the best. We further examine asymmetric models about the GC in order to see if the observed asymmetry of the spurs are better explained. 

\begin{figure} 
\begin{center} 
\includegraphics[width=8cm]{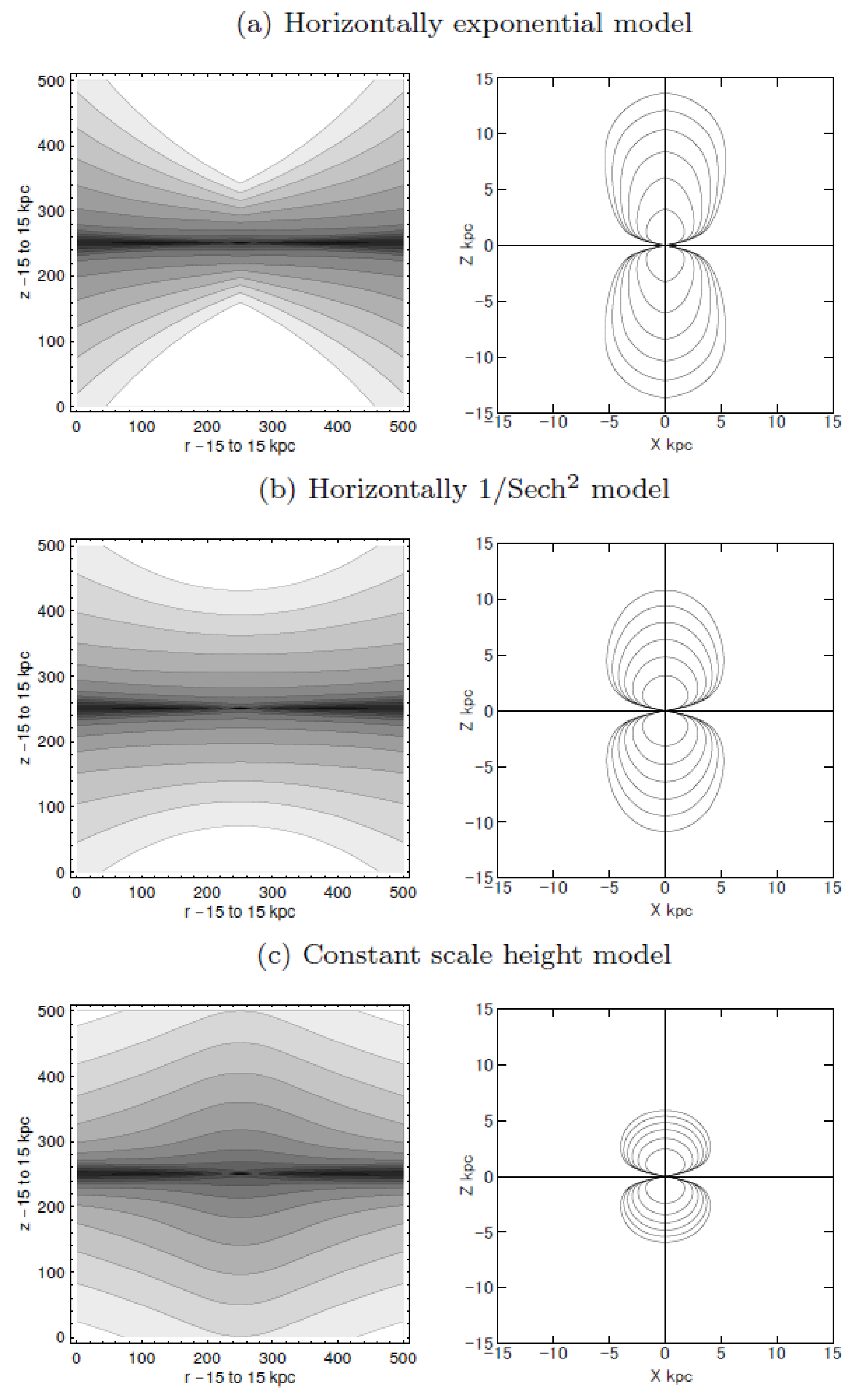} 
\end{center}
\caption{Meridional density maps of different halo models (left panels) and corresponding time evolution of the shock front every 2.5 My (right). (a) radially ($r$) exponential model,  (b) $\sech^2(z/z_i)$ model, and (c) constant scale height model. The contours in the density map are drawn every $\sqrt{10}$ times the neighbor starting from $10^{-5}$ (white region) and ending at 10 \Hcc (darkest region). }
 \label{comparison}  
\end{figure} 

{
\begin{table*}
\caption{Qualitative fitness of the calculated shock front to the observed GC spurs for different halo models}
\begin{tabular}{llll}
\hline\hline 
Model & Property & Shape on the sky & Fitness   \\
\hline  
Semi-exponential (standard) model& Variable scale heights & Elongated BHS$^\dagger$, fits to obs. & Good   \\
Exponential model  & ibid. Discontinuous at rot. axis & Edged shell top & Merginal \\
Cosh$^2$ model & ibid. Mild density profile. & Smaller sized BHS & Good, if $E$ is much larger. \\
Semi expo./Expo. model & Constant scale heights  & Too flattened BHS & Not good   \\
Spherical $\beta$ model & Politrope & Too flattened BHS & Not good   \\ 
 \hline
\end{tabular} \\ 
$^\dagger$ Bipolar hyper-shell.
\label{fitness}
\end{table*} 
}

\subsection{Spherical hot halo}

Besides the disk-like halo model, as adopted in the standard model, a spherical halo model so called $\beta$ model is often adopted (Miller and Bregman 2013), which is represented by
\be
\rho_{\rm X: \beta}=\rho_{\rm c}(1+(R/R_{\rm c})^2)^{-3\beta/2},
\label{sphhalo}
\ee
where $\rho_{\rm c}$ is the central density, $R_{\rm c}$ is the scale radius, or often called the core radius, and $\beta$ is the politropic index. By fitting the spherical halo model to soft X-ray observations, Nakashima et al. (2018) obtained $\rho_{\rm c}=1.2\times 10^{-3}$ \Hcc, $R_{\rm c}= 2.4$ kpc, and $\beta=0.51$. 

Figure \ref{model-sph} shows a calculated result for the same parameters as in figure \ref{model}, except that the X-ray halo density is given by equation \ref{sphhalo}. Because of the slower density decrease and, therefore, higher density in the outer halo than the exponentially decreasing profile, the shock front is strongly decelerated as it expands beyond a few kpc. The dumbbell shape becomes much flatter, and the observed round shape of the NPS and GC spurs are not well reproduced. Hence,it may be concluded that the hot halo may better be represented by the disk model. Furthermore, the spherical model predicts high density in the outer halos at $r>\sim 10$ kpc, extending far into the IGM till $\sim 30$ \kpc in the polar axis, which seems unrealistic in a normal-sized spiral galaxy.
 
\subsection{Halo with exponentially variable scale height}

Exponentially increasing scale height with the distance from the rotation axis is assume for the halo gas density. In this case the $\r$ profile has a singularity at $\r=0$ (rotation axis). As shown in figure \ref{comparison} (a), the density contours have discontinuity at the rotation axis. This results in the calculated front is slightly edged along the rotation axis. Except for this singularity, this model yields essentially the same result as the standard model.

\subsection{Halo with scale height varying as cosh$^2$}

Figure \ref{comparison}(b) shows a case for a halo with scale height increasing by cosh$^2$ function of the distance from the rotation axis. Because of the milder behavior of the scale height,  the halo gas density in the inner Galaxy is higher than that in the standard model. This yields less expanded front. However, this model gives essentially the same result as the standard model.

\subsection{Constant scale height model}

Figure \ref{comparison}(c) shows a constant-scale height model, where the scale heights $z_{\rm i}$ are taken constant over the Galactic disk, equal to the values at the normalized radii, respectively. The density profile is convex about the GC due to the slower decrease of th density with the height in the inner region than in the above models. Accordingly the halo density is higher in the inner Galaxy, and the shock front is less expanded compared to the standard model.

\section{Asymmetric halos: A qualitative consideration}

The observed GC spurs are asymmetric in morphology and intensity with respect to the Galactic plane and rotation axis (Sofue 1994, 2000; {\red Mou et al. 2018; Sarkar 2019}). It has been suggested that such asymmetry could be attributed to horizontal wind in the halo, or equivalently, motion of the Galaxy in the intergalactic gas  (Sofue 2000; {\red Mou et al. 2018}). The spurs also exhibit various inhomogeneous structures. We here examine how the asymmetry and inhomogeneity in the halo gas distribution affect the shock front propagation by modifying the standard model. However, the discussion will be only qualitative, because the fitting parameters are too many in order to reach a unique solution.

\begin{figure} 
\begin{center}  
\includegraphics[width=8cm]{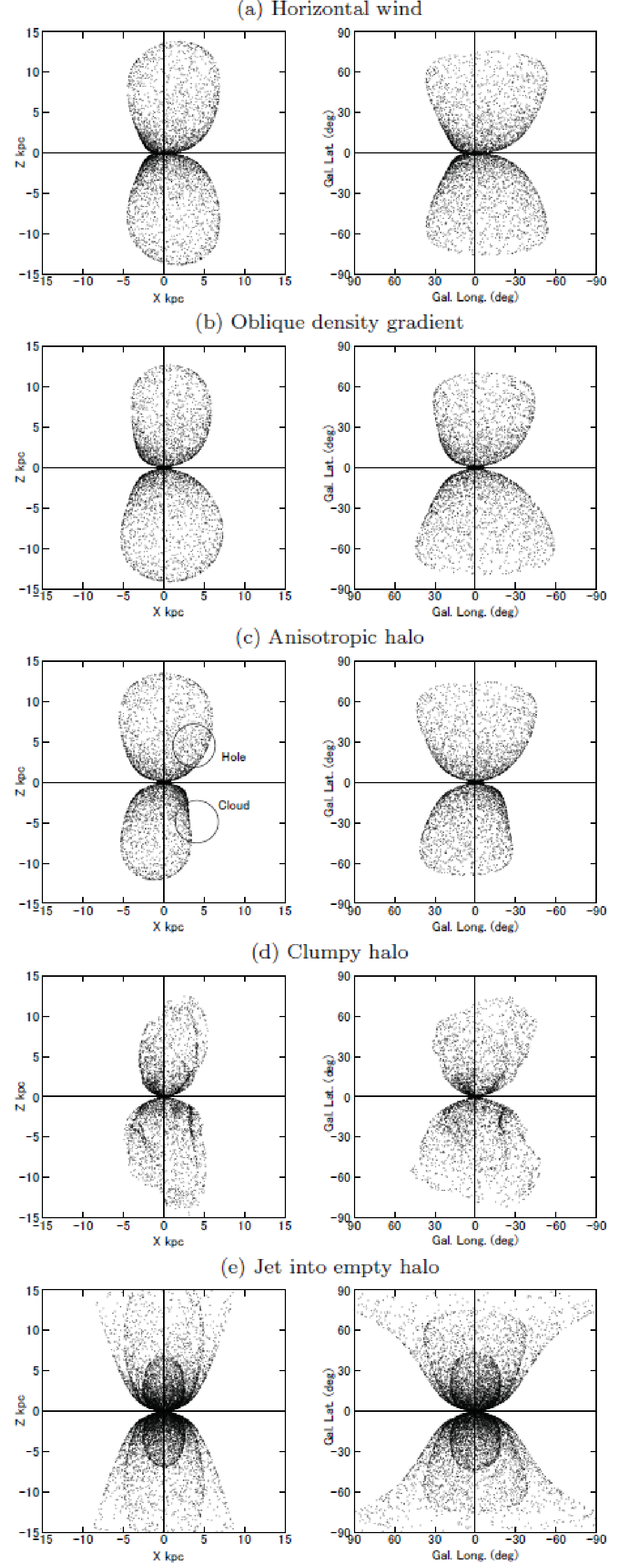}  
\end{center}   
\caption{Deformation of the BHS at $t=15$ My for $E=1.5\times 10^{55}$ erg expanding into a halo with (a) horizontal wind,
{(b) oblique density gradient from SE to NW, }
(c) anisotropy with large cavity and cloud, (d) sinusoidal clouds at arbitrary phases, and (e) truncated halo tops, forming a bipolar conical jet (5, 10 and 15 My). }
 \label{diag}   
\end{figure}

\subsection{Horizontal wind}

It has been suggested that the large-scale asymmetry of the NPS-E an NPS-W may be caused by a wind blowing in the halo (Sofue 2000; {\red Mou et al. 2018}).
In order to examine this idea, the shock propagation is calculated in the presence of an artificial ram pressure in the halo by adding a constant acceleration $g$ to the $X$-directional component of the second derivative of radius, $\ddot{R}$ of equation (\ref{eqsakashita}). 

Figure \ref{diag}(a) shows the result for acceleration represented by the ratio to the front acceleration, $g_{\rm w}=0.1\times$ and $0.2\times \ddot{R}_{\rm unit} \sim 3-6\times 10^{-7}$ cm s$^{-2}$, where $\ddot{R}_{\rm unit}=\ {\rm kpc}/1\ {\rm My^2}$ is the unit of the calculation. This may be compared with the acceleration of the front when NPS reaches $R\sim 8$ kpc at $\sim 8$ My, where $\ddot{R}\sim 4\times 10^{-7}$ cm s$^{-2}$.

Such an amount of acceleration is possible, if the ram pressure by the halo gas on the front increases by a factor of $g_{\rm w}/\ddot{R}\sim (V+V_{\rm wind})^2/V^2$, where $V$ and $V_{\rm wind}$ are the speeds of shell expansion and wind, respectively. Or, the wind velocity may be approximated by
$V_{\rm wind}\sim [(1+g/\ddot{R})^{1/2}-1]V$. 

From soft X-ray observations indicating hot plasma of $\sim 10^7$ K as well as from the present calculation, the expansion velocity is considered to be $V\sim 300$ \kms at the NPS. Then, the required wind velocity is estimated to be on the order of $V_{\rm wind}\sim 100-180$ \kms. 

The calculated result shows asymmetry of the front shape mimicking that of the NPS-E and W, and proves that a horizontal wind can indeed explain the observed asymmetry. Such a horizontal wind may be originated by motion of the Galaxy through the intergalactic space of the Local Group toward the east. This idea is consistent with the origin for the globally deformed HI disk having lopsided elongation (Nakanishi and Sofue 2016).
 
However, this wind model encounters a difficulty that the halo itself moves away from the Galaxy in a crossing time of the wind of $\sim 10\ \kpc/100$ \kms $\sim 10^8$ My. In order for the wind to blow steadily, the IGM gas must be constantly accreted and compressed in the flow, when the Galaxy is moving through the IGM at a subsonic velocity. However, it is beyond the scope of this paper to model such a wind mechanism. 
  
\subsection{Density gradient by IGM's ram pressure}

Another possible origin of the large-scale deformation of BHS is a density gradient across the entire halo. Such gradient may occur due to lopsided compression of the halo gas by the ram pressure of the IGM, when the Galaxy is moving through the Local Group. This model resembles the wind model, while it does not encounter the difficulty of halo stripping.

We assume that the halo stays stationary with the deformed shape unchanged, and that the gas density varies exponentially and obliquely across the galactic plane with horizontal and vertical scale lengths of variation of 3 and 6 kpc, respectively, giving a gradient from NE to SW direction on the sky. Figure \ref{diag}(b) shows the result at 15 My, which shows an east-west asymmetry as well north-south. 

{\red
In order for the halo to be deformed significantly, the IGM ram pressure is required to be comparable to the hot (X-ray) halo's internal pressure, which is comparable to the gravitational energy. This condition yields
$
\rho_{\rm IGM}V^2_{\rm IGM}\sim P,
$
where $P\sim n_{\rm X}kT$ is the pressure with $n_{\rm X}\sim 10^{-4}$ H cm$^{-3}$ and $kT \sim 0.2$ keV. For the IGM density of $n_{\rm IGM}\sim 10^{-5} $ H cm$^{-3}$, we obtain $V_{\rm IGM}\sim 10^3$ \kms, which is reasonable as random velocity of galaxies in the intergalactic space.  
}

\subsection{Large-scale irregularity}

Besides wind or large-scale density gradient in the halo, there may be various possibilities to produce asymmetry and deformation of the shock front such as due to density fluctuations, because the expansion velocity of the shock front is accelerated in a low-density region, and decelerated in a higher-density region. In fact significant density perturbations have been detected in the halo by differing sight line emission measures of the X ray emission (Gupta et al. 2014; 2017). 
{\red Sarkar (2019) has recently modeled a north-south asymmetry of the halo density, and showed an asymmetric expansion of a shock wave from the GC is indeed produced even by relatively small density perturbation.}

As one of such cases of inhomogeneous halo, we assume that the halo has a large clump and hole, or a cavity, of radius $\sim 3$ kpc, centered on $(X,Z)=(2,4)$ kpc and $(2,-4)$ kpc, whose central densities are 0.1 and 10 times the ambient density, respectively. 

Figure \ref{diag}(c) shows the result, where an asymmetric front shape is produced by faster expansion through the hole in the northern hemisphere, which seems to mimic the shape along NPS to NPS-W. On the other hand the wave through a cloud is deformed to form a locally concave front.

\subsection{Clumps and filaments}

Another case of inhomogeous halo is such that the halo is full of clouds with enhanced densities. Here, many clouds are added to the standard halo by multiplying the density profile by a factor $0.5+5.0 \Pi_{i=1,3} \cos^2[2\pi (x_i-a_i)/\lambda]$ in order to represent sinusoidal density fluctuations of wave length $\lambda/2$ and phases of the nodes of $a_i$. 

Figure \ref{diag}(d) shows an example for $\lambda/2=3$ kpc with arbitrary phases $a_i$. The global front shape remains similar to that of the standard model, while retardation of shock propagation through the clouds causes complicated deformation of the front. Side views of concave fronts produces filamentary features near the tangential projection of the whole front, resembling filaments in supernova remnant shells.

\subsection{Jet and/or wind into a truncated halo}

It may happen that the galactic halo is truncated by the interaction with the intergalactic gas such as due to the motion of the Galaxy through the IGM.
In order to calculate the shock wave propagation into a truncated halo, the density distribution for the standard model is multiplied by a factor of $\exp(-(z/z_{\rm h})^2)$, which represents a truncated halo beyond height $h$, where the density decreases lower than $\sim 10^{-5}$ \Hcc. Figure \ref{diag}(e) shows the result for $h=4$ kpc. 

The gas density decreases rapidly outward beyond the scale height, which causes acceleration of the front expansion in to the polar axis direction. The front attains a cylindrical jet or a conical wind in the outer halo along the polar axis. This is the mechanism for jet formation proposed by Sakashita (1971). Also, this model corresponds to the galactic wind model for the GC spurs proposed by Bland-Hawthorn and Cohen (2003).

The projection on the sky exhibits a widely open structure different from the observed loop shape near the polar axis, although the front appears still closed in so far as the shock front cone stays inside $|l|<90\deg$. 
It may be learnt that a conical wind or a jet of gas from the GC can be produced only if the halo is truncated at such low altitudes as $h\le \sim  4$ kpc. 
This implies in turn that a conical wind or a jet is difficult to be produced in the presence of a halo and IGM.

\begin{figure} 
\begin{center}       
\includegraphics[width=6cm]{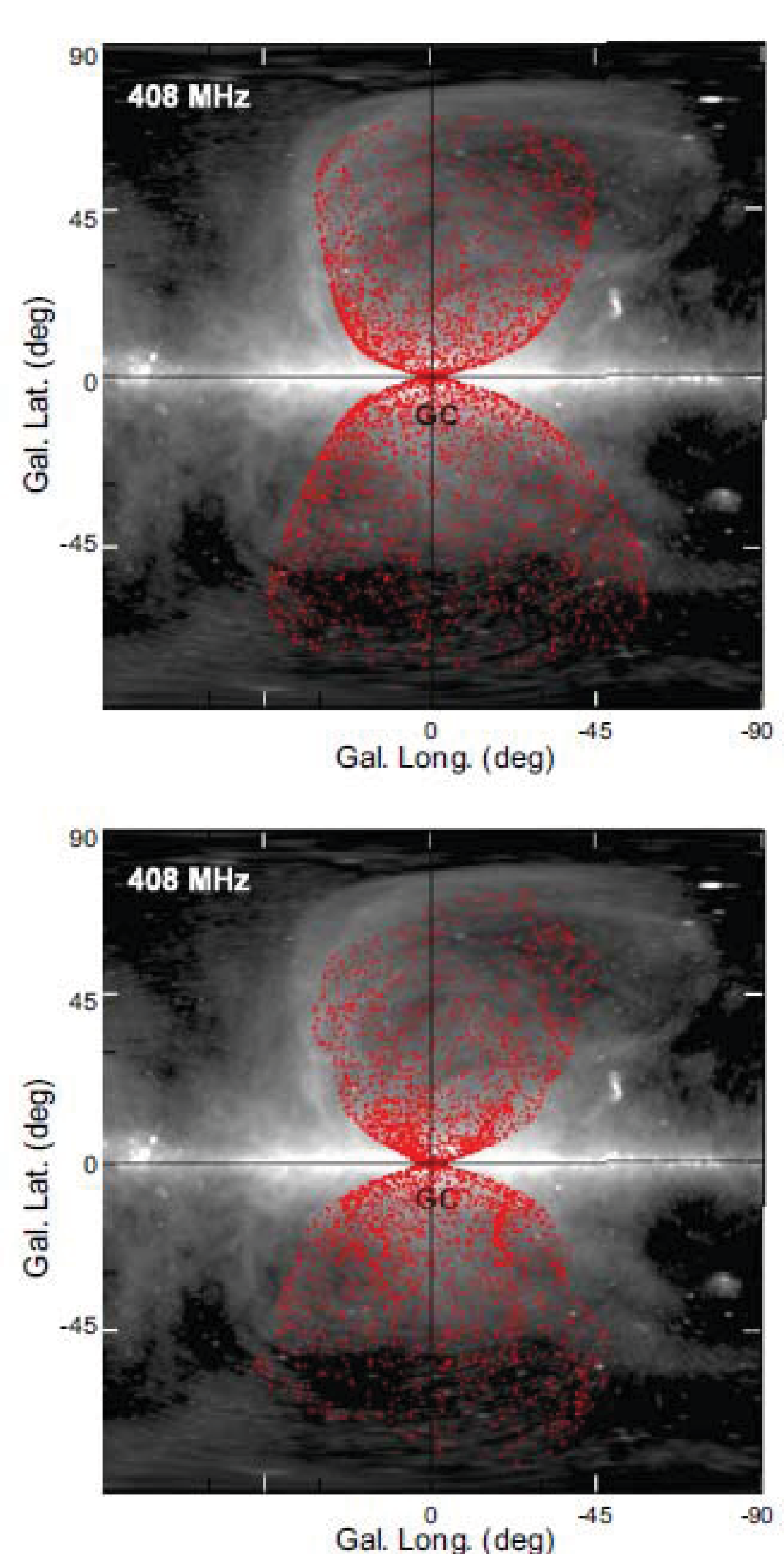}  
\end{center}  
\caption{[Top] Model front for a halo with oblique density gradient from top-left (north-east) to bottom-right (south-west), overlaid on the 408-MHz map. [Bottom] Same, but for a halo with inhomogeneous clouds.  }
 \label{overlay408}   
\end{figure}

{
\subsection{Overlay on 408 MHz map}

In figure \ref{overlay408} we overlay the calculated shock fronts for the density gradient model and sinusoidal clouds model on the 408 MHz continuum map in order to examine if the asymmetric spurs are explained by the models.  

It is found that any of the two models may be able explain the observed asymmetric features of the spurs, given appropriate parameters are chosen. It is, therefore, difficult to reach a unique solution about the density structure reponsible for the observed asymmetry of the spurs.
   }
   
\section{Summary and Discussion} 

The propagation of a shock wave produced by an explosive event at the GC such as due to a nucleus explosion or a starburst $\sim 15$ My ago with release energy $ 1.5\times 10^{55}$ erg was calculated by solving an improved Sakashita's (1971) equation taking into account the gravitational force of the Galaxy.
For solving the shock wave propagation, an advanced galactic disk+halo model was constructed based on HI, CO, UV and X-ray observations of the ISM in the Galaxy. 

The calculated shock front could reproduce the global morphology of the NPS and its counter spurs on the sky. It was found that the standard (semi-exponential) halo model appears best among the various models.  The spherical $\beta$-model was found to be not appropriate to reproduce the observed morphology of the GC spurs. 

{
The asymmetry of the spurs about the rotation axis (east-west asymmetry) and about the galactic plane (north-south asymmetry) is explained by the presence of a large-scale inhomogeneity in the halo. Both the model with large-scale density gradient across the entire Galaxy and the sinusoidal clouds model can qualitatively reproduce the observed asymmtric structures of the spurs.

A galactic wind in the halo may also be a cause of the asymmetry, while there remains difficulty about the stability of the halo itself. The density gradient or a wind in the halo may be caused by motion of the Galaxy through the IGM in the Local Group.

We also examined various kinds of models with inhomogeneous density distributions, which were shown to produce a variety of morphological properties of the spurs. Observed complicated structures superposed on the spurs such as filamentary features and/or dual/multiple curvatures can be qualitatively understood by combination of these inhomogeneous models.
  }
  
\section*{Acknowledgements}  
 The data analysis and computations were carried out on the data analysis computer system at the Astronomy Data Center of the National Astronomical Observatory of Japan.

\begin{appendix} 
 
\section{Sakashita's Method including Gravity}

{
We rewrite the equations of motion and energy conservation used in Sakashita's (1971) method by introducing gravitational energy, $\U$, and the difference between gravitational and centrifugal accelerations, $\G$, as follows (see Sakashita (1971) for details of the basic equations).
\be
{\d^2 r \/\d t^2}+{r^2\/\rho r} {\d P \/\d r}=g
\ee
and
\be
{E+\U \/ 4 \pi}=\int_0^R {P\/\gamma-1}r^2 dr+\int_0^R{1\/2}\({\d r \/\d t}\)^2\rho r^2 dr,
\ee
where $r$ and $t$ are the radius and time, $P$ is the total pressure including thermal and kinetic energy densities through an adiabatic relation between gas density and pressure, $\rho(r)$ is the ambient gas density in a fixed small cone. The gravitational forces are assumed to act in the radial direction. Although the gas near the galactic plane may be affected by vertical acceleration due to the flatness of disk, we neglect this effect, because it affects the front shape only near the galactic plane, and also because we are interested in the shock propagation in the halo.
}

Sakashita's (1971) equation for time variation of the front radius $R$ is, then, written as
{
$$
{E+\U \/4\pi}=
$$
}
$$
{J R\/3\gm}  \[{4(2\g-1)\/\gp^2}\ddot{R}+\G+{2\gm\/\gp^2}I \dot{R}^2+{8\gm \/\gp^3}{\dot{R}^2\/R}\] \\ 
$$
\be
+{2\/\gp^2}\dot{R}^2J + {2\/3\gm \gp}\rho R^3\dot{R}^2,
\label{eqsakashita} 
\ee
where $E$ is the explosion energy at the centre at $t=0$, $R$ is the radius of the shock front where all inside gas is accumulated, $\ddot{R}$ is the radial acceleration of the shock front, $g$ is the gravitational acceleration, $\theta$ is the elevation angle of the ray path measured from the galactic plane, so that the height from the plane is $z=r\ \sin \ \theta$, $\rho$ is the unperturbed density of the ambient gas at $(r,\theta)$, $\g=5/3$ is the adiabatic exponent, and
\be
\J=\int_0^R \rho(r,\theta)r^2dr={M (R,\theta) \/ 4 \pi}
\ee
is the mass inside $R$ divided by $4 \pi$ in a small cone at $\theta$, and
\be
\I=(d\ {\rm ln}\ \rho/dr)_R.
\ee 

\begin{figure} 
\begin{center}   
\includegraphics[width=6.5cm]{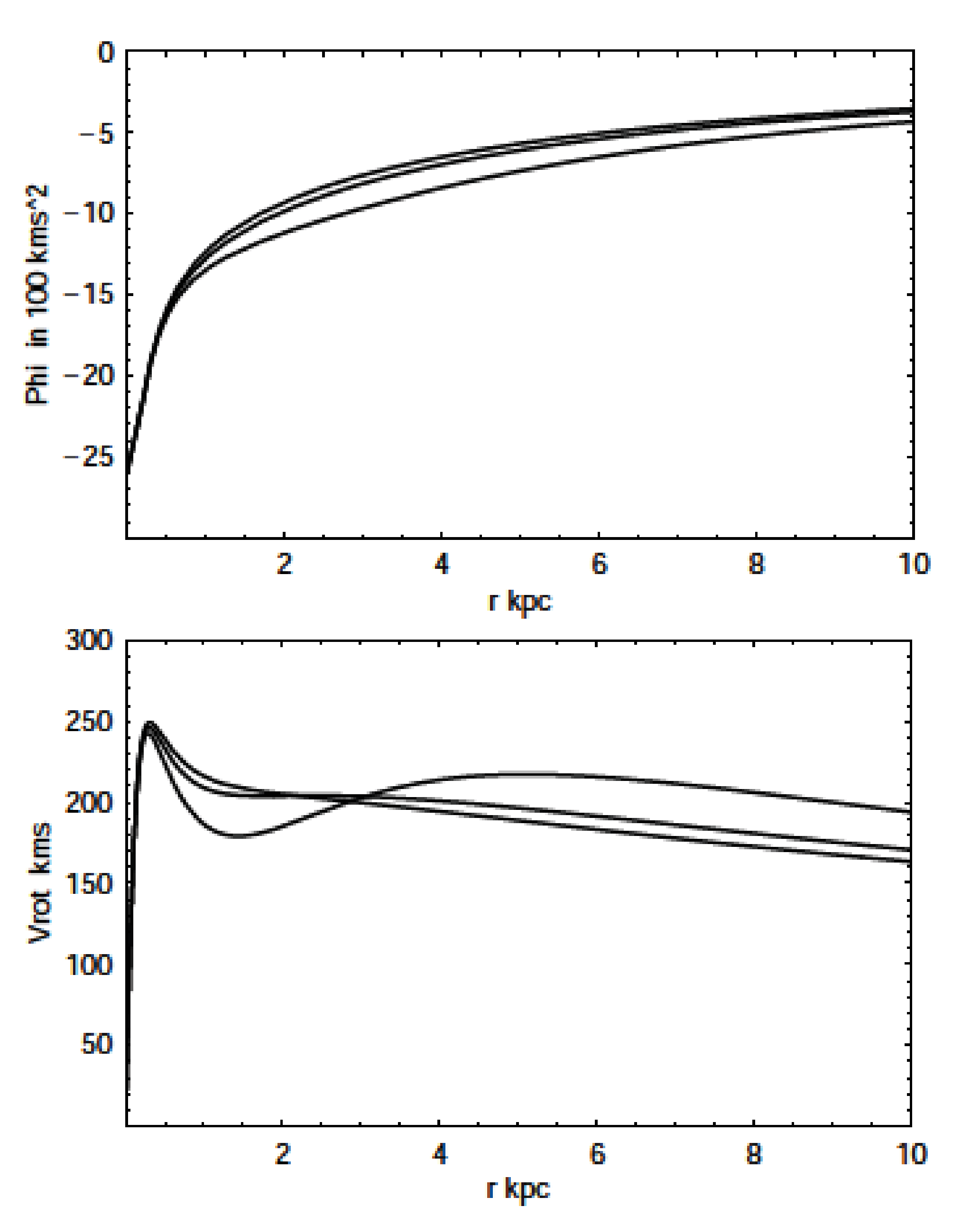} 
\end{center}
\caption{{[Top] Miyamoto-Nagai potential (1976) $Phi$ in \kms$^2$ at $\theta=0\deg, \ 45\deg$ and $90\deg$ from bottom line to top, respectively. [Bottom] Corresponding pseudo rotation curves from top to bottom. That at $\theta=0\deg$ represents the rotation curve in the disk.} }
\label{MNpotential}  
\end{figure}

{
The exhausted potential energy by the expansion against the gravity is given by 
\be
\U= \int_0^R [\Phi(r)-\Phi(R)] \rho(r) 4 \pi r^2 dr,
\ee
where $\Phi$ is the gravitational potential, and we adopt the Miyamoto-and-Nagai (MN: 1975) potential with the following parameters: $M_1=6\times 10^{9}\Msun, a_1=0$ and $b_1=0.2$ kpc for the bulge, and $M_2=10^{11}\Msun,\ a_2=3.5$ and $b_2=0.5$ kpc for the disk, where $M_i$ is the total mass and  $a_i$ and $b_i$ are horizontal and vertical scale radius, respectively, of the $i$-th mass component. Note that the original MN potential is represented in the $(\r, \ z)$ space with $\r=r\ \cos\ \theta$. Figure \ref{MNpotential} shows the potential at $\theta=0\deg,\ 45\deg$ and $90\deg$.  

The acceleration $g$ is given by the difference of gravitational and centrifugal accelerations by
\be
g=\({\d \Phi \/\d r}\)_R+{\Vav^2 \/R},
\label{g}
\ee
where the first term represents gradient of the potential, which is negative, and the second term is the centrifugal force due to rotation.
Here, $\Vav$ is pseudo rotation velocity of the snow-plowed shell at $R$ satisfying the angular momentum conservation,
\be
\Vav={\int_0^R \Vrot'(r) r \rho r^2 dr \/ R \int_0^R \rho r^2 dr}, 
\ee
where
\be
\Vrot'=\sqrt{r {\d \Phi \/ \d r}}
\ee
is a generalized rotation velocity including velocity dispersion, and leads to $\Vrot$ at $\theta=0$ in the galactic plane where the dispersion velocity is negligible. At high $\theta$ in the halo, equation (\ref{g}) may not be accurate. However, the gravity effect becomes negligible there, because the gravitational energy is much less than the input energy $E$ in the polar region due to the low density in the halo.

To summarize, the additional terms $\U$ and $\G$ are significant near the galactic plane, where the expansion velocity decreases rapidly due to dense gas disk, becoming comparable to the escaping (rotation) velocity. On the other hand, they are negligible in the polar region, where the shock velocity is much faster than the escaping velocity because of the low density and acceleration due to rapidly decreasing pressure toward outside. 
}

{
We comment that the adopted halo density model is not self-consistent in the sense that it may not satisfy the hydrostatic equilibrium in the presence of the gravitational field. However, since the time scale of the shock wave propagation is sufficiently shorter than the time scale of the gravitational deformation of the halo, we may consider that the assumption would yield reasonable approximation.
}

\section{Comparison with Sedov's Solution for Uniform Density.}

\begin{figure} 
\begin{center}    
\includegraphics[width=5.5cm]{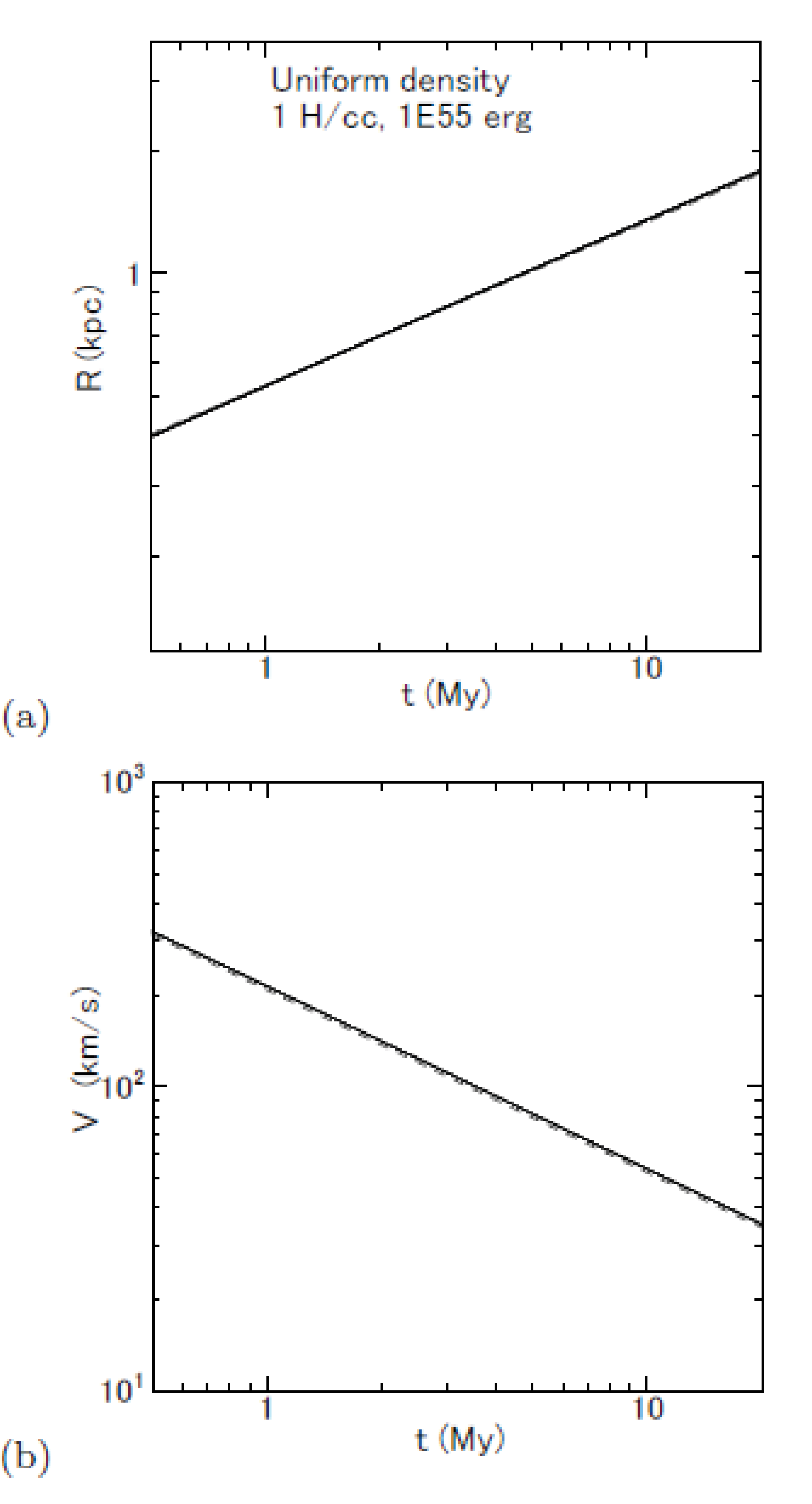}    
\end{center}
\caption{(a) Sedov solution (dashed line) compared with Sakashita's solution (solid line) for uniform density of $\rho_0=1$ \Hcc and explosion energy $E=10^{55}$ erg. (b) Same, but expansion velocity. 
}
 \label{Sedov}  
\end{figure}

The simplest solution for a uniform density gas without gravity is shown in figure \ref{Sedov} for $E=10^{55}$ erg and $\rho_0=1$ \Hcc in order to compare with the Sedov's similarity solution.

A simplified Sedov's similarity solution in an adiabatic uniform-density gas,
\begin{equation}
E={1\/2}M_0V^2={1\/2}{4\pi R^3\/3}\rho_0{\dot R}^2,
\end{equation}
is solved for the radius $R$, expansion velocity $V $ and elapsed time from the explosion $t$, yielding
\begin{equation}
V=\dot{R}=\(3\/2 \pi\)^{1\/2} \(E \/\rho_0\)^{1\/2}R^{-{3\/2}},
\end{equation}
\be
R=\(5\/2\)^{2\/3}\(3\/2\pi\)^{1\/5}\(E\/\rho_0 \)^{1\/5}t^{2\/5},
\ee
and
\be
V=\(5\/2\)^{-{3\/5}}\(3\/2 \pi \)\(E\/\rho_0\)^{1\/5}t^{-{3\/5}}.
\ee
The equations may be rewritten in unit of kpc, \kms, $10^{55}$ erg, and \Hcc as
\be
V=98.5 \ {\rm km\ s^{-1}}\ \(E^*\/\rho^*\)^{1\/2}\(R\/1{\rm kpc}\)^{-{3\/2}},
\ee
\begin{equation}
R=0.531 \  {\rm kpc}  \
\left( E^*\/ \rho^* \right)^{1\/5} 
\left( t\/ 1{\rm My}  \right) ^{2 \/5}
\end{equation}
and
\begin{equation}
V=208.8 \ {\rm km \ s^{-1}}  \ 
\left( E^*\/ \rho^*  \right)^{1\/5} 
\left( t\/ 1{\rm My} \right)^{-{3 \/5}},
\end{equation}
where $E^*=E/10^{55}$ erg, 
$\rho^*$ $=\rho_0/ 1$ \Hcc $=\rho_0/1.6735\times 10^{-24}$ $ {\rm g \ cm^{-3}}$. 

Figure \ref{Sedov} shows comparison of the results for these Sedov equations with those using the Sakashita's (1971) method calculated for an explosion of $E=10^{55}$ erg in a uniform density gas of $\rho_0=1$ \Hcc. Both results agree  with each other.

  {
\section{Standard Disk+Halo Model}  
\label{appdensitymodel}

In this appendix, we describe the details of individual gaseous components in the disk and halo for the standard model, which consists of (1) the central molecular zone (CMZ), (2) main molecular disk, (3) cold HI disk,  (4) warm HI halo (disk), (5) hot ionized (UV) halo (disk), (6) hot X-ray halo, and (7) uniform intergalactic gas (IGM). Each density is expressed in cylindrical coordinates $(\r, z)$ with $\r=r \ \cos \theta$ and $z=r \ \sin \theta$. 
Figure \ref{rhoprof} shows the density profiles in the $z$ direction at $r=3$ kpc and in the galactic plane. 
}

\subsection{Central molecular zone} 

The molecular disk is expressed by a superposition of the central molecular zone (CMZ) and main molecular disk. The CMZ is assumed to be expressed by a dense, thin Gaussian disk of molecular gas represented by
\be
\rho_1=\rho_{\rm CMZ}=\rho_{0,{\rm CMZ}} e^{-(\r/0.02\kpc)^2}e^{-(\r/0.3\kpc)^2}.
\ee
{
We assume $\rho_{0, {\rm CMZ}}=100$ \Hcc, and the total H$_2$ mass in the CMZ is $\sim 10^7\Msun$. 
}

\subsection{Main molecular disk}

\begin{figure} 
\begin{center}         
\includegraphics[width=7cm]{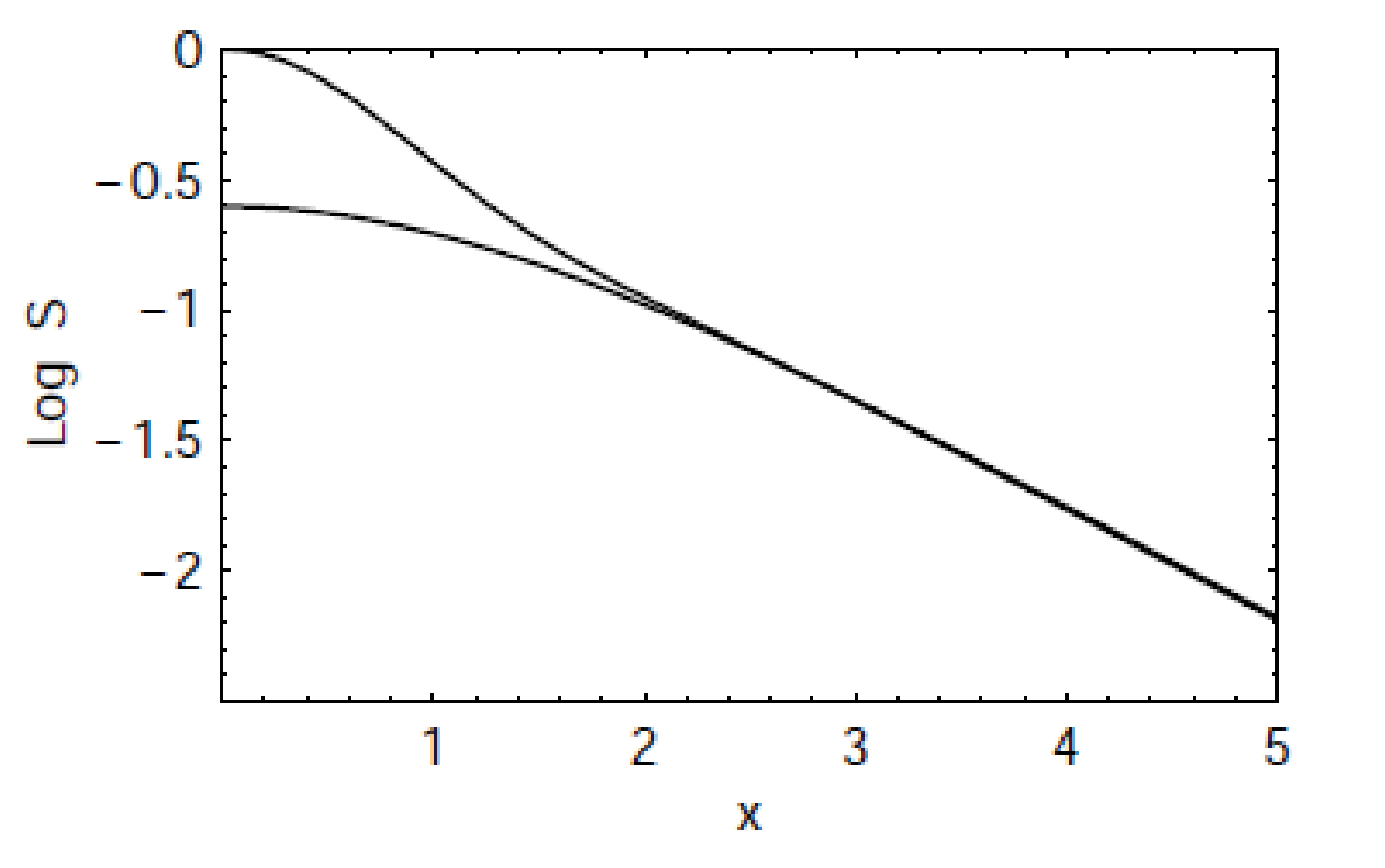} 
\end{center}
\caption{Comparison of semi-exponential function, $\S$, for $\alpha=1$ (upper line) with $(1/4) {\rm sech}^2(x/2)$ function (lower), both tending to exp($-x$) at $x\gg 1$.}
\label{SemiExpSech}  
\end{figure}

The  main molecular disk is represented by a $\sech^2$ function as 
\be
\rho_{\rm mol}=\rho_{\rm mol.0}\ \sech (z/z_{\rm mol})^2\A_{\rm mol}(\r),
\ee
where 
\be
z_{\rm mol}(\r)\sim 0.06\kpc \ \S(8{\rm kpc}/\ad)/\S(\r/\ad)
\ee
is the variable scale height inversely proportional to the $z$-directional gravity due to the Galactic stellar disk, whose scale radius is $\ad=5.73$ kpc (Sofue 2015b). 
In order to avoid discontinuity at the rotation axis,  a semi-exponential function, $\S(x)$, is introduced:
\be
\S(x)=f\ e^{-x}+(1-f)e^{-x^2},
\label{Sfunction}
\ee
where $f=x^2/(\alpha^2+x^2)$ is  the fractional function of the exponential and Gaussian components, so that $\S$ is approximated by Gaussian at $x\ll \alpha$ and tends to exponential at $x\gg \alpha$, and $\alpha=0.5$ is assumed here. Figure \ref{SemiExpSech} shows comparison of the $\S$ function with $(1/4) {\rm sech}^2(x/2)$ function.

The radial profile function $\A$ is defined by
\be
\A_{\rm mol}(\r)={0.06\kpc \/z_1}\B(\r,5 \kpc,4\kpc) \times \C(\r).
\ee
Here, the first term normalizes the surface density in the vertical direction to that at $r=3$ kpc, the second term $\B(\r)$ is a "ring function" representing a ring in the disk normalized at $r=3$ kpc defined by
\be
\B(\r,\r_{\rm ring},w)={e^{-((\r-r_{\rm ring})/w)^2}\/e^{-((3 \kpc-\r_{\rm ring})/w)^2}},
\ee
where, $r_{\rm ring}$ and $w$ are the ring radius and width. Note that the molecular disk is also well fitted by a Gaussian function, or $\sech^2(z/z_1)\simeq e^{-(z/z_1)^2}$ near the disk, which is similar to but different from those for the HI and warmer components as below.

The "hole function" $\C(\r)$ represents the central hole of the galactic gas disk inside 3 kpc, and is given by
\be
\C(\r)={(\r/3\kpc)^\eta \/ 1+(\r/3\kpc)^\eta }
\label{holefunc}
\ee
with $\eta = 2$. The hole function applies to the main molecular disk, HI, and warm HI disk. 

The central density $\rho_{\rm mol, 0}$ is normalized by the value at $r=3$ kpc (G22), which is estimated by applying the tangent-point method to the observed CO brithness temperature, 
\be  
\rho_{\rm mol, 0} \sim (1.36\times 2\mH/\sqrt{2})
X_{\rm CO} {\Tb}_{\rm CO} \sqrt{\sigma_{\rm mol} V_{\rm rot}} / R,
\label{rhoMol}
\ee
where $X_1=1\times 10^{20} {\rm H}_2 $  cm$^{-2}$ [\Kkms]$^{-1}$ is the CO-to-H{$_2$ column conversion factor in the inner Galaxy (Arimoto et al. 1993), $\sigma_{\rm mol}=5$ \kms is the velocity dispersion, $V_{\rm rot}=220$ \kms is the rotation velocity, and $r=3$ kpc. Inserting $\Tb\sim 2$ K as observed near G22, $\rho_{\rm mol}\sim 1$ \Hcc at $r\sim 3$ kpc.

\subsection{Cold HI disk}  
 
The cold HI component near the galactic plane is expressed as
\begin{equation}
\rho_{\rm HI}=\rho_{\rm HI, 0}  \sech^2\(z\/z_{\rm HI}\) \A_{\rm HI}(\r),
\end{equation} 
where $\rho_{\rm HI, 0}=0.4$ \Hcc,
\be
z_{\rm HI}(\r)\sim 0.2\kpc\ \S(8 {\rm kpc}/\ad)/\S(\r/\ad),
\ee
and 
\begin{equation}
\A_{\rm HI}(\r)={0.2\kpc \/z_{\rm HI}} \B(\r,10\kpc,8\kpc)\times \C(\r) .
\end{equation}
Here, the first term normalizes the surface density for varying scale height, $\B$ represents the large scale HI ring with radius $\sim 10$ kpc (Nakanishi and Sofue 2016), and $\C$ represents the HI hole inside 3 kpc.  
 
The coefficient was determined by measuring volume HI density of the galactic disk at $l\sim 22\deg-25\deg$ ($\r \sim 3$ kpc) by applying the tangent-point method (Sofue 2018) to the HI survey data by  McClure-Griffiths et al. (2009) as,
\begin{equation}
\rho_{\rm HI,0}=\rho_{\rm wHI}\sim ({\mH/\sqrt{2}})X_{\rm HI} {\Tb}_{\rm HI}{\sqrt{\sigma_{\rm HI} V_{\rm rot}} / \r},
\label{rhoHI}
\end{equation}
where $X_{\rm HI}=1.82\times 10^{18}$ H cm$^{-2}$ [\Kkms]$^{-1}$ is the conversion factors, $\sigma_{\rm HI}\sim 10$ is  velocity dispersion, $V_{\rm rot}=220$ \kms is the rotation velocity, and $\r= 3$ kpc is the galacto-centric distance of the tangent point.   
 
\subsection{Warm HI halo}  

Since the the extended HI and halo components are distributed mostly outside the stellar galactic disk, their $z$ profiles are approximated by the exponential function in a constant vertical gravitational field by a thin disk. The warm HI gas may be fitted by
\be
\rho_{\rm wHI}=\rho_{\rm warm HI}={\rho_{\rm wHI, 0z}} \exp(-|z|/z_{\rm wHI}) \A_{\rm wHI}(\r)
\ee 
with $\rho_3=0.02$ \Hcc, 
\be
z_{\rm wHI}(\r)\sim 0.4 \kpc \ \S(3 {\rm kpc}/\ad)/\S(\r/\ad) ,
\ee
and
\begin{equation}
\A_{\rm wHI}(\r)={0.4\kpc \/z_{\rm wHI}}\B(\r,10\kpc,8\kpc) \times \C(\r) 
\end{equation}
are factors similar to that for the cold HI.

\subsection{UV halo}

The classical hot halo (disk) of $\sim 10^5$ K and scale height of $\sim 0.44$ kpc has been known by absorption observations of UV lines such as CIV toward halo stars within several kpc near the Sun (Sembach and Savage 1992; Shull and Slavin 1994). Since the observations have been obtained in the solar neighborhood, while the shock wave propagation is calculated from GC to large radii and heights, the distribution is rewritten as follows. 
\be 
\rho_{\rm UV}=\rho_{\rm UV, 04} \exp(-|z|/z_4(\r)) \A_4(\r) ,
\ee
where 
\be
z_{\rm UV}(\r)=0.44 \kpc\ \S(R_0/\rdisk)/\S(\r/\rdisk)
\ee
is a variable scale height with the Galacto-centric distance $r$ moralized to the scale height $0.4$ kpc obtained near the Sun (Shull and Slavin 1994), 
$ \rdisk=5.73 $ kpc is the scale radius of the mass distribution in the Galactic disk (Sofue 2015).

The third term $\A_{\rm UV}(\r)$ normalizes the density at $r=R_0$ in order to approximate the observed X-ray halo molded by Nakashima et al. (2018), and to keep the vertically integrated surface density against the increasing scale height:  
\begin{eqnarray} 
\A_{\rm UV}(\r)={\S(\r/\r_{\rm UV}) \/ \S (R_0/\r_{\rm UV})} {z_{\rm UV}(R_0)\/z_{\rm UV}(\r)} \nonumber \\
={\S(\r/\r_{\rm UV}) \S(\r /\rdisk) \/ \S (R_0/\r_{\rm UV})\S(R_0/\rdisk)}.
\end{eqnarray}
The scale radius $r_{\rm UV}=\rdisk=5.73$ kpc is taken to be equal to that of the galactic disk's gravity from rotation curve fitting (Sofue 2015b). The density and scale height of the hot halo near the Sun ($R\sim R_0$) have been measured by Shull and Slavin (1994) to be $\rho_{\rm UV}= 0.01-0.02$ \Hcc and $z_{\rm UV}(R_0)=0.44$ kpc, while  $\rho_{\rm UV,0}=0.02$ \Hcc is adopted here.

\subsection{X-ray Halo} 

From soft X-ray observations, the hot halo has been found to be represented by an exponential flat disk of high-temperature gas of temperature $\sim 2\times 10^6$ K  multiplied by exponential radial profile (Nakashima et al. 2018). Modifying their result, the X-ray hot halo is represented by
\begin{equation} 
\rho_{\rm X}={\rho_{\rm X,0}} \exp(z/z_{\rm X}(\r)) \A_{\rm X}(\r)
\end{equation}  
where 
\be 
z_{\rm X}(\r)=z_{\rm X}(R_0) \ \S(R_0/\rdisk)/\S(\r/\rdisk),
\ee
with $\rho_{\rm X,0}=3.7\times 10^{-3}$ \mHcc, $z_{\rm X}(R_0)=1.8$ kpc, $\r_{\rm X}=\rdisk=5.73$ kpc, and  
\be
\A_{\rm X}(\r)={\S(\r /\r_{\rm X} ) \/ \S (R_0/\r_{\rm X})} {z_{\rm X}(R_0)\/z_{\rm X}(\r)}
={\S(\r/\r_{\rm X}) \S(\r/\rdisk) \/\S(R_0/\r_{\rm X}) \S(R_0/\rdisk )}.
\ee 
The scale radius is assumed to be equal to the disk's scale radius for simplicity of the model, while a larger value of 7 kpc has been obtained for the X-ray halo (Nakashima et al. 2018).

\subsection{Intergalactic gas} 

The intergalactic gas density is assumed to be uniform at $\rho_{\rm IGM}=10^{-5}$ \Hcc. \\

}
\end{appendix}


\begin{thebibliography}{}  

\bibitem[\protect\citeauthoryear{Akita et al.}{2018}]{2018ApJ...862...88A} Akita M., Kataoka J., Arimoto M., Sofue Y., Totani T., Inoue Y., Nakashima S., 2018, ApJ, 862, 88 


 
\bibitem[\protect\citeauthoryear{Berkhuijsen, Haslam, \& Salter}{1971}]{1971A&A....14..252B} Berkhuijsen E.~M., Haslam C.~G.~T., Salter C.~J., 1971, A\&A, 14, 252  

 
\bibitem[\protect\citeauthoryear{Crocker et al.}{2015}]{2015ApJ...808..107C} Crocker R.~M., Bicknell G.~V., Taylor A.~M., Carretti E., 2015, ApJ, 808, 107 


\bibitem[\protect\citeauthoryear{Dame, Hartmann, \& Thaddeus}{2001}]{2001ApJ...547..792D} Dame T.~M., Hartmann D., Thaddeus P., 2001, ApJ, 547, 792 


\bibitem[\protect\citeauthoryear{Fang, Bullock, \& Boylan-Kolchin}{2013}]{2013ApJ...762...20F} Fang T., Bullock J., Boylan-Kolchin M., 2013, ApJ, 762, 20 


\bibitem[\protect\citeauthoryear{Gupta et al.}{2014}]{2014Ap&SS.352..775G} Gupta A., Mathur S., Galeazzi M., Krongold Y., 2014, Ap\&SS, 352, 775 

\bibitem[\protect\citeauthoryear{Gupta, Mathur, \& Krongold}{2017}]{2017ApJ...836..243G} Gupta A., Mathur S., Krongold Y., 2017, ApJ, 836, 243 



\bibitem[Haslam et al.(1982)]{1982A&AS...47....1H} Haslam, C.~G.~T., Salter, C.~J., Stoffel, H., \& Wilson, W.~E.\ 1982, AAS, 47, 1    
 
  
\bibitem[Jonas et al.(1985)]{1985A&AS...62..105J} Jonas, J.~L., de Jager, G., \& Baart, E.~E.\ 1985, AA Suppl, 62, 105  
 

\bibitem[\protect\citeauthoryear{Kataoka et al.}{2018}]{2018Galax...6...27K} Kataoka J., Sofue Y., Inoue Y., Akita M., Nakashima S., Totani T., 2018, Galax, 6, 27 
  

\bibitem[\protect\citeauthoryear{McClure-Griffiths et al.}{2009}]{2009ApJS..181..398M} McClure-Griffiths N.~M., et al., 2009, ApJS, 181, 398 
 

\bibitem[\protect\citeauthoryear{Miller \& Bregman}{2013}]{2013ApJ...770..118M} Miller M.~J., Bregman J.~N., 2013, ApJ, 770, 118 

\bibitem[\protect\citeauthoryear{Miller \& Bregman}{2015}]{2015ApJ...800...14M} Miller M.~J., Bregman J.~N., 2015, ApJ, 800, 14 

\bibitem[\protect\citeauthoryear{Miller \& Bregman}{2016}]{2016ApJ...829....9M} Miller M.~J., Bregman J.~N., 2016, ApJ, 829, 9  

\bibitem[\protect\citeauthoryear{Miyamoto \& Nagai}{1975}]{1975PASJ...27..533M} Miyamoto M., Nagai R., 1975, PASJ, 27, 533 

\bibitem[\protect\citeauthoryear{Moellenhoff}{1976}]{1976A&A....50..105M} Moellenhoff C., 1976, A\&A, 50, 105 

\bibitem[\protect\citeauthoryear{Mou, Sun, \& Xie}{2018}]{2018ApJ...869L..20M} Mou G., Sun D., Xie F., 2018, ApJ, 869, L20  

\bibitem[\protect\citeauthoryear{Nakanishi \& Sofue}{2016}]{2016PASJ...68....5N} Nakanishi H., Sofue Y., 2016, PASJ, 68, 5  
 
\bibitem[\protect\citeauthoryear{Reich, Testori, \& Reich}{2001}]{2001A&A...376..861R} Reich P., Testori J.~C., Reich W., 2001, A\&A, 376, 861  

\bibitem[\protect\citeauthoryear{Sakashita}{1971}]{1971Ap&SS..14..431S} Sakashita S., 1971, ApSS, 14, 431   

\bibitem[Sarkar et al.(2015)]{2015MNRAS.453.3827S} Sarkar, K.~C., Nath, B.~B., \& Sharma, P.\ 2015, MNRAS, 453, 3827 

\bibitem[\protect\citeauthoryear{Sarkar et al.}{2016}]{2016ApJ...818L..24S} Sarkar K.~C., Nath B.~B., Sharma P., Shchekinov Y., 2016, ApJ, 818, L24 

\bibitem[\protect\citeauthoryear{Sarkar}{2019}]{2019MNRAS.482.4813S} Sarkar K.~C., 2019, MNRAS, 482, 4813  

\bibitem[\protect\citeauthoryear{Shchekinov}{2018}]{2018Galax...6...62S} Shchekinov Y., 2018, Galax, 6, 62  

\bibitem[\protect\citeauthoryear{Shull \& Slavin}{1994}]{1994ApJ...427..784S} Shull J.~M., Slavin J.~D., 1994, ApJ, 427, 784 

\bibitem[\protect\citeauthoryear{Sembach \& Savage}{1992}]{1992ApJS...83..147S} Sembach K.~R., Savage B.~D., 1992, ApJS, 83, 147  

\bibitem[\protect\citeauthoryear{Snowden et al.}{1997}]{1997ApJ...485..125S} Snowden S.~L., et al., 1997, ApJ, 485, 125   
   
\bibitem[\protect\citeauthoryear{Sofue}{1977}]{ } Sofue, Y. 1977, AA 60, 327. 

\bibitem[\protect\citeauthoryear{Sofue}{1984}]{1984PASJ...36..539S} Sofue Y., 1984, PASJ, 36, 539   
  
\bibitem[\protect\citeauthoryear{Sofue}{}]{ }Sofue, Y.  1994, ApJ.L., 431, L91
  
\bibitem[\protect\citeauthoryear{Sofue}{2000}]{2000ApJ...540..224S} Sofue Y., 2000, ApJ, 540, 224 

\bibitem[\protect\citeauthoryear{Sofue}{2015}]{2015PASJ...67...75S} Sofue Y., 2015, PASJ, 67, 75 

\bibitem[\protect\citeauthoryear{Sofue}{2017}]{2017PASJ...69L...8S} Sofue Y., 2017, PASJ, 69, L8 

\bibitem[\protect\citeauthoryear{Sofue}{2018}]{2018PASJ...70...50S} Sofue Y., 2018, PASJ, 70, 50  

\bibitem[\protect\citeauthoryear{Sofue et al.}{2016}]{2016MNRAS.459..108S} Sofue Y., Habe A., Kataoka J., Totani T., Inoue Y., Nakashima S., Matsui H., Akita M., 2016, MNRAS, 459, 108    

\bibitem[Su et al.(2010)]{2010ApJ...724.1044S} Su, M., Slatyer, T.~R., and Finkbeiner, D.~P.\ 2010, ApJ, 724, 1044       

\end{thebibliography}
\end{document}